# Missing and spurious interaction in additive, multiplicative and odds ratio models

Fernandez-de-Cossio, J.* †, Fernandez-de-Cossio-Diaz, J. ‡, Takao, T.¥, Perera, Y. §
†Bioinformatics Department, §Molecular Oncology Group, Pharmaceutical Division, Center for Genetic Engineering and Biotechnology, PO Box 6162, CP10600, Havana (Cuba).
‡Systems Biology Department, Center of Molecular Immunology, PO Box 6162, CP10600, Havana (Cuba).
¥Laboratory of Protein Profiling and Functional Proteomics, Institute for Protein Research, Osaka University, Suita 565-0871, Japan
*jorge.cossio@cigb.edu.cu

## ABSTRACT

Additive, multiplicative, and odd ratio neutral models for interactions are for long advocated and controversial in epidemiology. We show here that these commonly advocated models are biased, leading to spurious interactions, and missing true interactions.

## INTRODUCTION

The concept of interaction stands for long controversial in epidemiology circles [1]–[7]. Main stream of these approaches arrive to additive effects as a meaningful model for testing "biologic interaction" [8][9][5], and deny, except in few special cases [10], a similar role for multiplicative effects. The issue is problematic because case-only designs for gene-environment and gene-gene interactions rely on the validity of multiplicative models [11][12], and have been regarded as more efficient alternatives to the typical random sampling of cases and controls [13].

We bring two hypothetical examples *a priori* detached from any explicit mathematical equation, but amenable to be unambiguously classified as interaction and non-interaction, respectively. We here evaluate the models for interaction dominant in epidemiology circles, by pursuing at least consistency with these two instances of our notion of interaction.

## INTERACTION NOTION

Colloquially, by interaction we mean that the concurrent exposition to a set of factors produces an effect that is surprising when compared to what one would expect from the effects of the intervening factors, if they acted independently.

For example, spark and oxygen (factors) are each practically innocuous, with only the spark releasing a small and transient amount of heat (the effect). In the presence of gasoline, inner factors "unavailable" to the analysis can be regarded as the particular chemical configuration holding the potential energy that is released only in the join exposition to spark and oxygen that lead to the surprising effect (fire) compared to their separated individual effects. Internally, the chemical bonds storing the stabilizing energy in gasoline break out by oxidation with $O_2$ (combustion), and the chemical (organic compound) changes into $CO_2$, $H_2O$, generating heat, light, sound, as the effect of an exothermic reaction (Figure 1).

Spark, oxygen and gasoline as interaction in the effect example

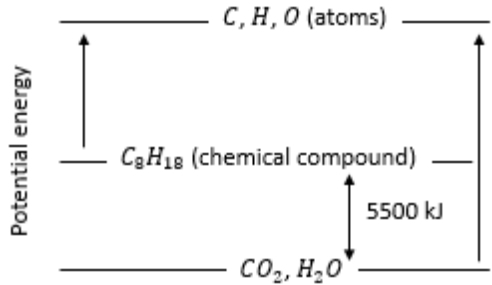

*Figure 1: Interactions in an exothermic reaction*

# INTERACTION SCENARIOS

We adopt from [14] two hypothetical examples *a priori* detached from any explicit mathematical equation, but amenable to be unambiguously classified as interaction and non-interaction, respectively.

Briefly, suppose the effect $E$ is manifested in three variants $E_A$, $E_B$ and $E_C$, corresponding to exposition to factors $A$, $B$ and $C$ respectively. The situation can be expressed by $E = E_A \vee E_B \vee E_C$, where ∨ denote the logical Or.

### No-interaction example
Each factor has independent pathways toward the effect, that is, a sequence of no intersecting events lead to the effect. For an illustrative example see Figure 1 in [14].

### Interaction example
In the interaction cases there is a cross-over of the pathways. For an illustrative example see Figure 1 in [14]. Factor $C$ activate a sequence of events that yield effect $E_C$, except for two indispensable process 1 and 2. These two process are each switched-on only by exposition to factors variants $a \in A$ and $b \in B$ respectively. Exposition to $a$ alone activate process 1 in the path from $C$ toward the effect, but the process 2 remain inactive impeding the completeness of the pathway, hence no effect is caused by factor $C$. The same arise in the lower path. Only the simultaneous exposition to $a \in A$ and $b \in B$ allows the concurrent activation of both process 1 and 2, completing the process sequence yielding to effect $E_C$. As a consequence, besides the individual effects $E_A$ and $E_B$ produced by $a$ and $b$, there is an additional effect $E_C$ not shared by the individual factors, that contribute to the total effect $E$.

# QUANTITATIVE FRAMEWORK

To gain a quantitative understanding, we introduce a notation for the probabilities of occurrence of the effects under different exposition to the factors.

Let a binary factor $A$ triggers events leading to $E_A$ with probabilities $\Pr(E_A|a) = p_a$ and $\Pr(E_A|\bar{a}) = p_{\bar{a}}$. The overbar denotes logical Not or negation (ex. $\bar{a}$ denotes no exposition to factor $A$). A binary factor $B$ trigger causal events leading to $E_B$ with probabilities $\Pr(E_B|b) = p_b$ and $\Pr(E_B|\bar{b}) = p_{\bar{b}}$.

We can assume that $\Pr(\bar{E}_A|A\,B) = \Pr(\bar{E}_A|A)$, $\Pr(\bar{E}_B|\bar{E}_A\,A\,B) = \Pr(\bar{E}_B|B)$, and $\Pr(\bar{E}_C|\bar{E}_A\bar{E}_B\,A\,B) = \Pr(\bar{E}_C|A\,B)$. By the rules of probability theory we have for every combination instances of factor $A\,B$ that

$$\Pr(\overline{E}|A\,B) = \Pr(\overline{E}_A|A)\Pr(\overline{E}_B|B)\Pr(\overline{E}_C|A\,B) \qquad (1)$$

In the interaction case, when factor $C$ is active, the join exposition to $a \in A$ and $b \in B$ activates both process 1 and 2 yielding effect $E_C$ with probability $\Pr(E_C|a\,b) = p_c$. Let the probability at any other combination of factors $A$ and $B$ be equal, i.e. $\Pr(E_C|a\,\overline{b}) = \Pr(E_C|\overline{a}\,b) = \Pr(E_C|\overline{a}\,\overline{b}) = p_{\overline{c}}$. To show the bias in the models, it is sufficient to find one or more instances demonstrating the inconsistency. For the derivations that follows it is more simple to let $p_{\overline{c}} = 0$. In this case $E_C$ is an impossible event if no interaction arise, then $E_C$ uniquely represents an occurrence of $E$ due to the interaction between $A$ and $B$, and only in this case $p_c > 0$.

Replacing ( 1 ) with the particular factor instances yields

$$\Pr(\overline{E}|a\,b) = (1 - p_a)(1 - p_b)(1 - p_c)$$
$$\Pr(\overline{E}|a\,\overline{b}) = (1 - p_a)(1 - p_{\overline{b}})$$
$$\Pr(\overline{E}|\overline{a}\,b) = (1 - p_{\overline{a}})(1 - p_b) \qquad (2)$$
$$\Pr(\overline{E}|\overline{a}\,\overline{b}) = (1 - p_{\overline{a}})(1 - p_{\overline{b}})$$

Summarizing, no-interaction require $p_c = 0$, and the model for interaction require that $p_c > 0$.

## BIAS IN MODELS OF INTERACTIONS

Now we can demonstrate that the commonly advocated models for interaction, additive, multiplicative of risk ratios and odds ratios, are biased. The effect measure typically adopted in epidemiology is the relative risk ratio $RR$, defined by

$$RR_{ij} = \frac{\Pr(E|i\,j)}{\Pr(E|\overline{a}\,\overline{b})}$$

Where $\Pr(E|i\,j)$ is the probability of the effect $E$ under exposition to factors $i \in \{a, \overline{a}\}$ and $j \in \{b, \overline{b}\}$.

### Additivity of risk ratios

The additive model relies on this equality for no-interaction

$$1 = \frac{RR_{ab} - 1}{RR_{\overline{a}b} + RR_{a\overline{b}} - 2}$$

By rearranging we arrive to this equality

$$\Pr(\overline{E}|a\,b) + \Pr(\overline{E}|\overline{a}\,\overline{b}) = \Pr(\overline{E}|\overline{a}\,b) + \Pr(\overline{E}|a\,\overline{b}) \qquad (3)$$

We will demonstrate that this equality can be satisfied for $p_c > 0$, contradicting what is supposed to occur in the non-interaction scenario. Replacing according to ( 2 )

$$(1 - p_a)(1 - p_b)(1 - p_c) + (1 - p_{\overline{a}})(1 - p_{\overline{b}}) = (1 - p_{\overline{a}})(1 - p_b) + (1 - p_a)(1 - p_{\overline{b}})$$

Rearranging

$$1 - p_c = \frac{(1 - p_{\bar{a}})(1 - p_b) + (p_{\bar{a}} - p_a)(1 - p_{\bar{b}})}{(1 - p_a)(1 - p_b)}$$

We note that $p_c > 0$ when the right side is less than one, and this take place if

$$(p_a - p_{\bar{a}})(p_b - p_{\bar{b}}) < 0$$

By taking $p_{\bar{a}} > p_a$ and $p_b < p_{\bar{b}}$ or vice versa we can find infinite assignments of $p_a$, $p_{\bar{a}}$, $p_b$ and $p_{\bar{b}}$ yielding $0 < p_c$, spanning the whole range $p_c \in (0,1)$. Therefore, the additive model equality for no interaction can be satisfied for the interaction scenario, showing inconsistency.

We will now demonstrate the opposite, that equality ( 3 ) can be violated for $p_c = 0$, contradicting what is supposed to occur under interaction. Replacing ( 3 ) according to ( 2 ) and rearranging

$$(1 - p_a)(1 - p_b) + (1 - p_{\bar{a}})(1 - p_{\bar{b}}) = (1 - p_{\bar{a}})(1 - p_b) + (1 - p_a)(1 - p_{\bar{b}})$$

The above equality is not satisfied for $p_a = p_b = 1 - x$, $p_{\bar{a}} = p_{\bar{b}} = 1 - y$ where $0 < x < y < 1$, since $x^2 + y^2 = 2xy$ only for $(x - y)^2 = 0$, that is for $x = y$.

Hence, the additive model can miss actual interactions (false negative), and introduce spurious interactions (false positive).

### Multiplicativity of risk ratios

This model [3][12] rely on the neutrality function for no-interaction defined by this equality

$$1 = \frac{RR_{ab}}{RR_{\bar{a}b} \, RR_{a\bar{b}}}$$

By rearranging we arrive to the equality

$$\Pr(E|a\,b) \Pr(E|\bar{a}\,\bar{b}) = \Pr(E|\bar{a}\,b) \Pr(E|a\,\bar{b}) \qquad (4)$$

Replacing according to ( 2 )

$$\{1 - (1 - p_a)(1 - p_b)(1 - p_c)\} \times \{1 - (1 - p_{\bar{a}})(1 - p_{\bar{b}})\}$$
$$= \{1 - (1 - p_{\bar{a}})(1 - p_b)\} \times \{1 - (1 - p_a)(1 - p_{\bar{b}})\}$$

If we replace for example $p_a = 0.5$, $p_b = 0.3$, $p_{\bar{a}} = 0.2$, we can span the whole range $0 < p_c < 1$ of infinite solutions satisfying the multiplicative rule by assigning

$$p_c = \frac{8.079951555341193 - 26.93317185113728 \, p_{\bar{b}}}{6.284406765265366 + 25.137627061061472 \, p_{\bar{b}}}$$

for any values $0 < p_{\bar{b}} < 0.333$, as ploted in Figure 2.

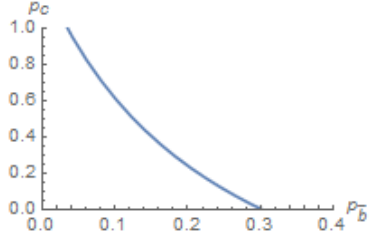

Figure 2: $p_c$ span the whole range $0 < p_c < 1$ when $p_{\bar{b}}$ vary from 0 to 0.333.

To demonstrate that equality ( 3 ) can be violated for $p_c = 0$, replace ( 4 ) according to ( 2 ) and rearranging

$$\{1 - (1 - p_a)(1 - p_b)\} \times \{1 - (1 - p_{\bar{a}})(1 - p_{\bar{b}})\}$$
$$= \{1 - (1 - p_{\bar{a}})(1 - p_b)\} \times \{1 - (1 - p_a)(1 - p_{\bar{b}})\}$$

The above equality is not satisfied for $p_a = p_b = 1 - x$, $p_{\bar{a}} = p_{\bar{b}} = 1 - y$ where $0 < x < y < 1$.

Hence, the multiplicative model can miss actual interactions (false negative), and introduce spurious interactions (false positive).

### Odds ratios

Multiplicativity of odds ratios is also commonly advocated in epidemiology circles [11][13]. This model relies on the equality

$$\frac{\Pr(E|a\,b)}{\Pr(\bar{E}|a\,b)} = \frac{\Pr(E|\bar{a}\,b)}{\Pr(\bar{E}|\bar{a}\,b)} \frac{\Pr(E|a\,\bar{b})}{\Pr(\bar{E}|a\,\bar{b})} \Big/ \frac{\Pr(E|\bar{a}\,\bar{b})}{\Pr(\bar{E}|\bar{a}\,\bar{b})} \qquad (5)$$

By rearranging equation ( 5 ), we arrive to the equality

$$\frac{\Pr(E|a\,b)\Pr(E|\bar{a}\,\bar{b})}{\Pr(\bar{E}|a\,b)\Pr(\bar{E}|\bar{a}\,\bar{b})} = \frac{\Pr(E|\bar{a}\,b)\Pr(E|a\,\bar{b})}{\Pr(\bar{E}|\bar{a}\,b)\Pr(\bar{E}|a\,\bar{b})}$$

Replacing with ( 2 ) and after expansions and simplifications yields

$$1 - p_c = \frac{p_{\bar{a}} + p_{\bar{b}} - p_{\bar{a}} p_{\bar{b}}}{p_a p_b + p_{\bar{a}} + p_{\bar{b}} - p_a p_{\bar{b}} - p_{\bar{a}} p_b}$$

The condition $p_c = 0$ is only satisfied when

$$p_a(p_b - p_{\bar{b}}) = p_{\bar{a}}(p_b - p_{\bar{b}})$$

Or equivalently when $p_a = p_{\bar{a}}$ or $p_b = p_{\bar{b}}$. Figure 3 show the values spanned by $p_c$ when the no-interaction criteria is satisfied under the multiplicative odds ratios model.

To demonstrate that Odds ratios equality can be violated for $p_c = 0$, rearrange

$$1 = \frac{p_{\bar{a}} + p_{\bar{b}} - p_{\bar{a}} p_{\bar{b}}}{p_a p_b + p_{\bar{a}} + p_{\bar{b}} - p_a p_{\bar{b}} - p_{\bar{a}} p_b}$$

To obtain

$$p_{\bar{a}}(p_{\bar{b}} + p_b) = p_a(p_b - p_{\bar{b}})$$

Let $p_b = p_{\bar{b}} > 0$, and the above equality is violated for all values of $p_{\bar{a}} > 0$.

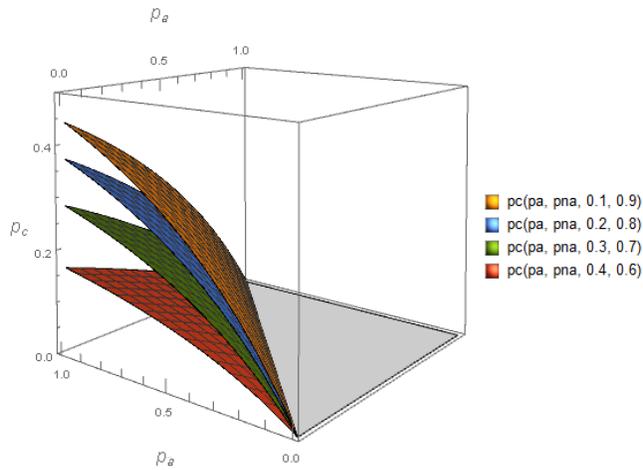

Figure 3: Plots of the values spanned by $p_c$ in the example scenarios, along different values of $p_a$, $p_b$, $p_{\bar{a}}$ and $p_{\bar{b}}$

Hence, the odd ratios model can miss actual interactions (false negative), and introduce spurious interactions (false positive).

# CONCLUSION

From the advent of case only designs, proponents emphasised that rationality of these approaches rely on the validity of multiplicative models of risk ratios and odds ratios. We demonstrated here that multiplicative risk ratio and odds ratio models are biased. Therefore, case-only designs are weakly funded in general.

Nowadays, huge amount of data are produced at unprecedented rates by high-throughput technologies, and put available in public repositories. Analysis of these data is challenging, and visualization and analysis tools are rapidly coping the web to deals with these rich information data. Commonly advocated additive, multiplicative and odds risk ratio models of interaction are implemented within these tools, and should be used with caution in epidemiology data.

# ACKNOWLEDGMENTS

This work has been partially supported by the Center for Genetic Engineering and Biotechnology, and performed in part under the International Cooperative Research Program of Institute for Protein Research, Osaka University, ICR-17-05.

# COMPETING FINANCIAL INTERESTS

The authors declare no competing financial interests